\documentclass[12pt]{article}
\usepackage{graphicx}
\usepackage{dcolumn}
\usepackage{amsmath,amssymb}
\usepackage{bm}

\begin{document}

\title{Natural majorization of the Quantum Fourier Transformation
in phase-estimation algorithms}

\author{Rom\'an Or\'us$^{\dag}$,
Jos\'e I. Latorre$^{\dag}$
and Miguel A. Mart\'{\i}n-Delgado$^{\ddag}$
\\
\\ \emph{$^{\dag}$Dept. d'Estructura i Constituents de la Mat\`eria,}
\\ \emph{Univ. Barcelona, 08028. Barcelona, Spain.}
\\ \emph{$^{\ddag}$Departamento de F\'{\i}sica Te\'orica I,}
\\ \emph{Universidad Complutense, 28040. Madrid, Spain.}
}
\maketitle

\begin{abstract}

We prove that  majorization relations hold step by step in the
Quantum Fourier Transformation (QFT) for phase-estimation algorithms. Our result
relies on the fact that states which are
mixed by Hadamard operators at any stage of the computation
only differ by a phase. This property is a consequence of
the structure of the initial state and of the QFT, based on controlled-phase
operators and a single action of a Hadamard gate per qubit.
The detail of our proof shows that
Hadamard gates sort the probability
distribution associated to
the quantum state, whereas controlled-phase operators carry all
the entanglement but are immaterial to majorization.
We also prove that majorization in  phase-estimation algorithms
follows in a most natural way from unitary evolution, unlike its
counterpart in Grover's algorithm.

\end{abstract}


\newpage

\section{Introduction}

Majorization theory emerges as the natural framework
to analyse and quantify  the measure of disorder for
classical probability distributions \cite{muirhead} \cite{hardy} \cite{marshall} \cite{maj}. Majorization
ordering is far more severe than the one proposed by standard
Shannon entropy. If one probability distribution
majorizes another, a set of inequalities must hold
that  constrain the former probabilities with respect to the latter.
These inequalities entail entropy ordering, but the
converse is not necessarily true.
Quantum mechanically, majorization has proven to be
at the heart of the solution of a large number
of quantum information problems.  It has been shown
that majorization plays a fundamental role in topics like 
ensemble realization, conversion of quantum
states via local operations and classical
communication, and characterization of positive operator valued
measurements \cite{vid}.

Recently \cite{lat}, a Majorization Principle has been formulated
and checked in the efficient quantum algorithms known so far:
the family of Grover algorithms and the family of quantum-phase
estimation algorithms.
More precisely,
Grover's \cite{grov} and Shor's \cite{shor} algorithms
operate in such a way that the  probability distribution associated to the
quantum state in the computational basis is majorized step by step
 until it reaches an optimal state. This property
has been completely proven in the case of Grover's
algorithm for search of an item in an unstructured database.
The algorithm is based on the iterative application of a
unitary transformation. It is easy to see \cite{lat}  that
each of these transformations majorizes the probability
distribution associated to the computational basis provided
the initial state is symmetric.
This step-by-step majorization progresses smoothly
until the algorithm reaches the solution
state after ${\cal O}(\sqrt{N})$ operations, where $N$ is the number of
entries in the database.

The case of phase-estimation algorithms, which include Shor's
factorisation algorithm, Simon's algorithm, clock synchronisation and
Kitaev's algorithm for the Abelian stabiliser problem
\cite{disc} is more subtle. The key ingredients in all these algorithms
are the use of the
quantum Fourier transformation operator (QFT) and the promise of a
specific structure of the initial state.
In \cite{lat} it was checked that the
canonical form of the QFT majorizes
 step by step the probability distribution attached
to the computational basis.
Here we provide a complete proof of how the notion of
majorization formulated in \cite{lat} explicitly operates in the
special case of phase-estimation quantum algorithms, and how this
notion is related to the efficency shown by these algorithms.

One of our main purposes is to present an explicit and detailed
proof of the following proposition:
\emph{Majorization works step by step in the
QFT of phase-estimation algorithms}. The whole property is based on the following ideas: Hadamard operators act majorizing the probability distribution given the symmetry of the quantum state, and such a symmetry is partially preserved under the action of both Hadamard and controlled-phase gates \cite{lat}.
The mathematical formulation of these concepts will lead us to define an ``H($i$)-pair'' as that pair of computational states that can be mixed by a Hadamard operator acting over the $i$-th qubit. Furthermore,  we also
work out a property concerning the way in which
majorization emerges naturally from unitary evolution for this type
of quantum algorithms.

In order to assess the  significance of step by step majorization we
have analyzed a series of further examples
that will be presented in a separate publication \cite{inpreparation}.
 First, we have analyzed a variant of
Berstein-Vazirani  algorithm \cite{bv} where a quantum algorithm is
able to obtain the slope of a linear function defined on
$\mathbb{Z}_N$ 
using only one query to the function. The efficient algorithm
solving this problem is of interest because no entanglement is ever
present along the computation, while majorization is verified.
Second, we have considered the problem of determining the parity
of a given function using oracle calls as introduced in ref.
\cite{parity}. The quantum oracular algorithm proposed  showed no speed-up as well 
as no step-by-step majorization. Third, we have studied the set of quantum adiabatic
algorithms proposed by Farhi et al. The efficiency of these algorithms
has been studied in detail for Grover's problem showing that
non-linear evolutions do lead to more efficient algorithms \cite{adi}. We have
indeed checked that the efficient time-path is associated to step-by-step
majorization
while the non-efficient time-path is not. Finally, we have
studied the recently proposed quantum random walk algorithm to
solve a classicaly hard graph problem \cite{randomwalks}.
This algorithm also shows a step-by-step minorization-majorization cycle.
These four instances of step-by-step majorization will be
presented elsewhere but can be considered here as motivation
for a  link between majorization and
efficiency.

 We have structured the paper as follows. In Sec. 2 we
review some concepts about majorization theory and how
they relate to quantum algorithms. We develop in Sec. 3
some  properties of majorization in  phase-estimation
algorithms and present in detail
 the main problem to be solved.  In Sec. 4 we
produce the proof for QFT step-by-step majorization. In Sec. 5 we
analyse the way in which majorization
arises in these algorithms and, finally,  in Sec. 6 we collect
 our conclusions.

\section{Majorization theory and its relation to quantum
algorithms}

Let us review the notion of majorization formulated in \cite{lat}
for quantum algorithms.
Let us consider two vectors $\vec{x}$, $\vec{y}\in
\mathbf{R}^d$ such that $\sum_{i=1}^d x_i = \sum_{i=1}^d y_i = 1$,
whose components represent two different probability
distributions . We say
that  distribution $\vec{y}$ majorizes distribution $\vec{x}$ (written as
$\vec{x}\prec \vec{y}$) if, and only if, there is a set of
probabilities $p_j$ and permutation matrices $P_j$ such that
\begin{equation}
\vec{x} = \sum_j p_j P_j \ \vec{y} \ .
\label{maj1}
\end{equation}
This definition gives us the intuitive notion that the $\vec{x}$
distribution is more disordered than $\vec{y}$, because the former
can be obtained from the latter making a probabilistically weighted
 sum over
permutations of $\vec{y}$. There is an alternative definition of
majorization which is often more practical.
Consider  the  components of the two vectors sorted in decreasing
order, written as $(z_1, \ldots z_d) \equiv
\vec{z}^\downarrow$. We say that $\vec{y}^\downarrow$ majorizes
$\vec{x}^\downarrow$ if, and only if, the following set of inequalities holds:
\begin{equation}
\sum_{i=1}^k x_i \leq \sum_{i=1}^k y_i \qquad k = 1 \ldots d-1 \ .
\label{maj2}
\end{equation}
A third way of defining majorization involves the use of doubly stochastic matrices. A real $d \times d$ matrix $D=(D_{ij})$ is said to be doubly stochastic if its entries are non-negative, and each row and column of $D$ sums to 1. We say that $\vec{y}$ majorizes $\vec{x}$ if, and only if, there exists a doubly stochastic matrix $D$ such that
\begin{equation}
\vec{x} = D \vec{y} \ .
\label{maj3}
\end{equation}
The three given definitions can be proven to be equivalent \cite{maj}.

Majorization can be related to quantum algorithms in the following
way: let $|\psi_m \rangle $ be the pure state representing the
register in a quantum computer at an operating stage labelled by
$m = 0, 1\ldots M-1$, where $M$ is the total number of steps of
the algorithm. We can naturally associate a set of sorted
probabilities $p_{x}$, $ x = 0, 1\ldots 2^n -1$ to this quantum
state of $n$ qubits in the following way: decompose the register
state in the computational basis, i.e:
\begin{equation}
|\psi ^{(m)}\rangle = \sum_{x=0}^{2^n-1}c_{x}^{(m)}|x\rangle \ ,
\label{maj4}
\end{equation}
where $\{|x\rangle \equiv |x_{n-1}, x_{n-2} \ldots x_{0} \rangle
\}_{x=0}^{2^n -1}$ denotes the basis states in binary notation,
and $x = \sum_{j=0}^{n-1}x_j2^j$. The probability distribution
associated to this state is
\begin{equation}
\label{maj5}
\vec p^{(m)}=\{p_x^{(m)}\}\qquad
 p_{x}^{(m)} \equiv |c_{x}^{(m)}|^2 = |\langle x | \psi^{(m)} \rangle
|^2 \qquad  x = 0, 1 \ldots 2^n - 1\ ,
\end{equation}
corresponding to the set of probabilities to find each
possible output. A quantum algorithm will be said to majorize
step by step this probability distribution iff \cite{lat}
\begin{equation}
\label{maj6}
\vec p^{(m)} \prec \vec p^{(m+1)} \qquad \forall
m=1,\dots,M \ .
\end{equation}
In such a case, there will be a neat flow of probability directed
to the values of highest weight, in a way that the probability
distribution will be steeper and steeper as the time arrow goes
on.

It is important to note that majorization is attached to a
probability distribution defined on a specific basis. Although
majorization is basis dependent, the basis where the final measurement
closing the algorithm is performed is singled out. The computational
basis is often the natural measurement basis to analyze majorization since it
gives  the  probability distribution associated to an eventual
measurement. We could stop the computation at some
arbitrary time and find, if step-by-step majorization is holding,
that the probability distribution is orderly approaching
the final one. There may be instances, like the determination
of parity \cite{parity}, where the measurement basis is
different from the naive computational basis. The
relevant concept of majorization should then be analyzed in
the physical measurement basis.

\section{Majorization in phase-estimation
quantum algorithms}

Phase-estimation algorithms, initially introduced by Kitaev \cite{kit}, 
form a family of efficient quantum
algorithms \cite{disc} that include, for instance,
 Shor's factoring algorithm and discrete logarithms \cite{shor}. 
Their relevance is due to the
exponential gain in computational time over known classical
algorithms. The basic problem to solve can be stated as follows.
Given an unitary operator $U$ and one of its eigenvectors
$|v\rangle$, estimate the phase of the corresponding eigenvalue
$U|v\rangle = e^{-2\pi i \phi}|v\rangle$, $\phi \in [0,1)$ with
$n$ qubits accuracy. An efficient solution was found in
\cite{disc} and can be summarised in the following series of steps,
represented by the quantum circuit of Fig. 1:

(i) Prepare the pure state $|\psi^{(i)} \rangle = |00 \ldots \rangle |v
\rangle$, where $|00\ldots\rangle$ will be called the register state
and $|v \rangle$ is the source state  where we have stored
the eigenvector of the unitary operator $U$.

(ii) Apply Hadamard operators
\begin{equation}
U_H =\frac{1}{\surd2}\left( \sigma_1 + \sigma_3 \right)
\label{had}
\end{equation}
over all the qubits in the register state.

(iii) Apply bit-wise controlled $U^j$ gates over the $|v\rangle$ state
as shown in the Fig 1, where each $U^j$ gate corresponds to
the application of $j$ times the proposed $U$ gate with $j = 0, 1\ldots n-1$.

\bigskip

\begin{picture}(400,220)(0,210)
\put(0,395){$|0\rangle$} \put(15,398){\line(1,0){10}}
\put(25,386){\framebox(24,24){$U_H$}}
\put(49,398){\line(1,0){151}} \put(205,395){$|0\rangle + e^{-2 \pi
i \phi 2^{n-1}} |1\rangle $} \put(300,398){\line(1,0){10}}
\put(340,398){\line(1,0){15}}

\put(0,359){$|0\rangle$} \put(15,362){\line(1,0){10}}
\put(25,350){\framebox(24,24){$U_H$}}
\put(49,362){\line(1,0){151}} \put(205,359){$|0\rangle + e^{-2 \pi
i \phi 2^{n-2}} |1\rangle $} \put(300,362){\line(1,0){10}}
\put(340,362){\line(1,0){15}}

\put(37,335){$\vdots$} \put(100,335){$\vdots$}
\put(250,335){$\vdots$} \put(355,335){$\vdots$}

\put(0,315){$|0\rangle$} \put(15,318){\line(1,0){10}}
\put(25,306){\framebox(24,24){$U_H$}}
\put(49,318){\line(1,0){151}} \put(205,315){$|0\rangle + e^{-2 \pi
i \phi 2^{0}} |1\rangle $} \put(300,318){\line(1,0){10}}
\put(340,318){\line(1,0){15}}

\put(0,265){$|v\rangle$} \put(15,268){\line(1,0){40}}
\put(55,252){\framebox(32,32){$U^{2^0}$}}
\put(87,268){\line(1,0){30}}
\put(117,252){\framebox(32,32){$U^{2^{n-2}}$}}
\put(149,268){\line(1,0){15}}
\put(164,252){\framebox(32,32){$U^{2^{n-1}}$}}
\put(196,268){\line(1,0){145}} \put(97,275){$\cdots$}
\put(97,257){$\ldots$} \put(345,265){$|v\rangle$}
\put(310,295){\framebox(30,126){$QFT$}}

\put(68.5,315){$\bullet$} \put(71,284){\line(0,1){32}}
\put(130.5,359){$\bullet$} \put(133,284){\line(0,1){76}}
\put(177.5,395){$\bullet$} \put(180,287){\line(0,1){109}}

\put(350,403){$\downarrow$} \put(350,367){$\downarrow$}
\put(350,323){$\downarrow$}

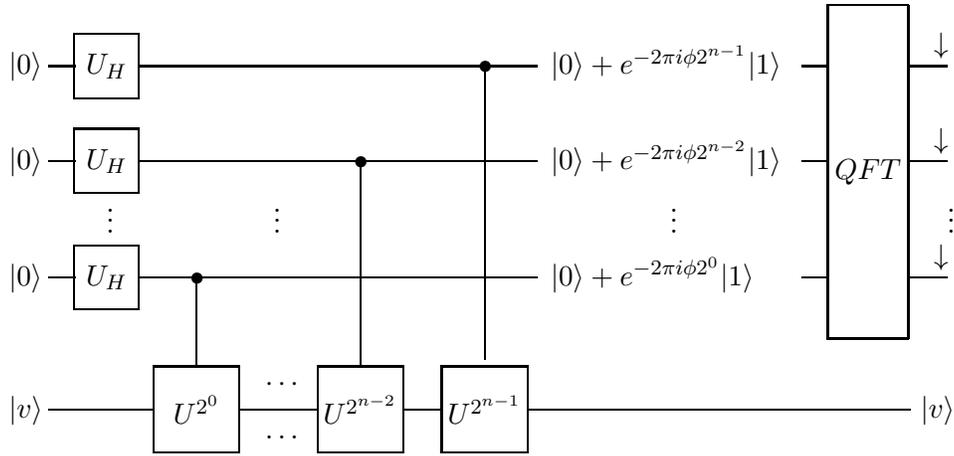
\put(0,225){Fig. 1: Quantum circuit corresponding to the
phase-estimation algorithm.}
\put(0,210){Arrows at the end
indicate measurements.}

\end{picture}

\bigskip
\bigskip
\bigskip

\begin{picture}(400,230)(0,-230)

\put(0,-40){\line(1,0){10}} \put(10,-52){\framebox(24,24){$U_H$}}
\put(34,-40){\line(1,0){5}} \put(39,-52){\framebox(24,24){$U_2$}}
\put(63,-40){\line(1,0){20}} \put(83,-52){\framebox(24,24){$U_n$}}
\put(67,-35){$\ldots$} \put(67,-47){$\ldots$}
\put(107,-40){\line(1,0){248}}

\put(0,-74){\line(1,0){112}}
\put(112,-86){\framebox(24,24){$U_H$}}
\put(136,-74){\line(1,0){5}}
\put(141,-86){\framebox(24,24){$U_2$}}
\put(165,-74){\line(1,0){20}}
\put(185,-86){\framebox(24,24){$U_{n-1}$}} \put(169,-69){$\ldots$}
\put(169,-81){$\ldots$} \put(209,-74){\line(1,0){146}}

\put(0,-108){\line(1,0){355}} \put(222,-103){$\ldots$}
\put(222,-115){$\ldots$}

\put(0,-150){\line(1,0){249}}
\put(249,-162){\framebox(24,24){$U_H$}}
\put(273,-150){\line(1,0){5}}
\put(278,-162){\framebox(24,24){$U_2$}}
\put(302,-150){\line(1,0){53}}

\put(0,-184){\line(1,0){307}}
\put(307,-196){\framebox(24,24){$U_H$}}
\put(331,-184){\line(1,0){24}}

\put(30,-140){$\vdots$} \put(150,-140){$\vdots$}
\put(345,-140){$\vdots$}

\put(48.5,-77){$\bullet$} \put(51,-52){\line(0,-1){22}}
\put(92.5,-187){$\bullet$} \put(95,-52){\line(0,-1){133}}
\put(150.5,-111){$\bullet$} \put(153,-86){\line(0,-1){22}}
\put(195.5,-187){$\bullet$} \put(198,-86){\line(0,-1){99}}
\put(286.5,-187){$\bullet$} \put(289,-162){\line(0,-1){24}}

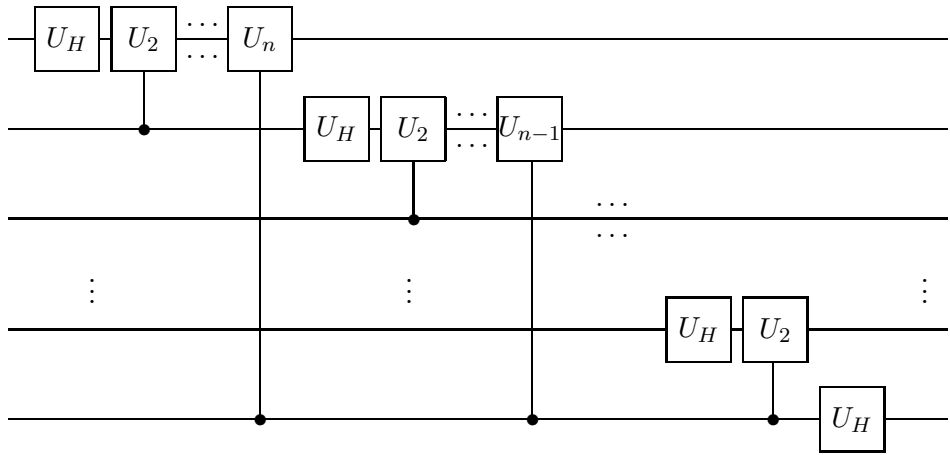
\put(0,-215){Fig. 2: Canonical decomposition of the QFT
operator. By $U_j$ we denote } \put(0,-230){the controlled gate
$|0\rangle \langle 0| + e^{2 \pi i /2^j}|1\rangle \langle 1 |$. }
\end{picture}

\newpage

(iv) Apply the  QFT operator
\begin{equation}
 QFT:|q\rangle \rightarrow
\frac{1}{2^{n/2}}\sum_{q'=0}^{2^n-1}e^{2\pi i q q'/2^n}|q'\rangle
\label{fourier}
\end{equation}
over the register state.

(v) Make a measurement of the register state of the system. This
provides with high probability the corresponding eigenvalue of $U$
with the required precision.

Let us now go through the algorithm focusing on how the
majorization of the computational basis probabilities
evolves.
The application of the Hadamard gates in step {\it (ii)} to the initial state
 produces a lowest element of majorization,
\begin{equation}
|\psi^{(ii)} \rangle = 2^{-n/2}\sum_{x=0}^{2^n-1}|x \rangle |v
\rangle \ ,
\label{stepdos}
\end{equation}
yielding to the probability distribution $p_x^{(ii)} = 2^{-n} \ \forall
x$. The outcome of the controlled $U^j$ gates in step {\it (iii)} is
the product state
\begin{eqnarray}
\label{steptres}
& |\psi^{(iii)} \rangle = 2^{-n/2}\left(|0\rangle
+ e^{-2 \pi i 2^{n-1}
 \phi}|1\rangle \right)\cdots \left(|0\rangle + e^{-2 \pi i 2^0 \phi}|1\rangle \right)|v\rangle
   \nonumber \\
& = 2^{-n/2}\sum_{x=0}^{2^n-1}e^{-2 \pi i x \phi }|x\rangle
|v\rangle \ .
\end{eqnarray}
As the action of these gates only adds local phases in
 the computational basis, the uniform distribution for the
 probabilities
is maintained ($p_x^{(iii)}=2^{-n} \ \forall x$).

Verifying majorization for the global action of the QFT is straightforward.
After step {\it (iv)} the quantum state becomes
\begin{equation}
 |\psi^{(iv)} \rangle = 2^{-n}\sum_{x,y=0}^{2^n-1}e^{-2 \pi i x (\phi - y/2^n) }|y\rangle |v\rangle \ .
\label{stepcuatro}
\end{equation}
We then have the probability distribution
\begin{equation}
  p_{y}^{(iv)}=\left|2^{-n} \sum_{x=0}^{2^n-1}e^{-2 \pi i x (\phi -
  y/2^n)}\right|^2 \qquad \forall y\ .
\label{finalprob}
\end{equation}
Majorization between the initial {\it (ii)} and final {\it (iv)}
states is verified \cite{lat},
according to the definition in (\ref{maj2}).
The remaining step {\it (v)} corresponds to a measurement whose
output is controlled with the probability distribution
$p_y^{(iv)}$.

The quantum speed up in the phase-estimation algorithm
is rooted in the efficient processing of the QFT.
It is then essential to investigate whether the
majorization property so far observed is also present
within the QFT step by step.

\section{Step-by-step majorization of the Quantum Fourier Transformation in the phase-estimation algorithm}

The mathematical statement about QFT we need to prove
reads:

\vspace{8pt}

\textbf{Theorem}

\emph{The QFT  majorizes step by step the
probability distribution calculated in the computational basis
as used in the phase-estimation algorithms.}

\vspace{8pt}

{\bf } This theorem is seen to emerge from two facts. It is,
first, essential that the initial state entering the QFT has
a certain symmetry to be discussed. Second, the order of the
action of Hadamard and controlled-phase gates maintains as much of
this symmetry as to be used by the rest of the algorithm. More
precisely, Hadamard gates take the role of majorizing the
probability distribution if some relative phases are properly
protected. Controlled-phase transformations do preserve such a
symmetry.

We  divide the proof in three steps: the first one will consist on
a majorization lemma (here is where majorization enters), the
second one will consist on a phase preserving lemma, and finally
the third one will be the analysis of the controlled-phase
operators in the QFT.  As hinted above, we will observe that
the only relevant operators for the majorization procedure are the
Hadamard gates acting over the different qubits, while
controlled-phase operators, thought providing entanglement, will
turn out to be immaterial to majorization.

\subsection{A lemma concerning majorization}

Let us introduce the concept of  ``H($i$)-pair'', central to
this paper. Consider a Hadamard gate  $U_{H,i}$ acting on qubit $i$ of
the quantum register. In general, the quantum register would
correspond to a superposition of states. This superposition
can be organized in pairs, each pair being characterized
by the fact that the Hadamard operation on qubit $i$ will mix
the two states in the pair. Let us illustrate
this definition with the example of a general quantum state of two qubits:
\begin{eqnarray}
|\psi \rangle &&  =  \alpha|00\rangle + \beta|01\rangle + \gamma|10\rangle + \delta|11\rangle  \nonumber \\
&& = \underbrace{ (\alpha|00\rangle + \beta|01\rangle)}_{{\rm H(0)-pair}}+ \underbrace{ (\gamma|10\rangle + \delta|11\rangle )}_{{\rm H(0)-pair}} \nonumber \\
&& = \underbrace{ (\alpha|00\rangle + \gamma|10\rangle)}_{{\rm H(1)-pair}} + \underbrace{(\beta|01\rangle + \delta|11\rangle)}_{{\rm H(1)-pair}} \ .
\label{pairs}
\end{eqnarray}
The second line corresponds to 
organizing the state as $H(0)$-pairs, because each pair
shares the first qubit value. The third line, instead, 
organizes the state on $H(1)$-pairs, because each
pair shares the second qubit value.

We can now formulate the following
lemma:

\vspace{8pt}
\textbf{Majorization lemma}

\emph{Let $|\psi \rangle$ be a pure
quantum state,
 written in the computational basis as}
\begin{equation}
|\psi \rangle = \sum_{x = 0}^{2^n-1}c_x |x \rangle \ ,
\label{state}
\end{equation}
\emph{with the property that the probability amplitudes of the
computational H($i$)-pairs differ only by a phase for a
given qubit $i$.}

\emph{Then, the probability distribution resulting from
$U_{H,i}|\psi \rangle$ majorizes the one resulting from $|\psi \rangle$.}

\vspace{8pt}

{\bf Proof}

Consider a $n$-qubit pure quantum state $|\psi \rangle$
with the assumed property that the probability amplitudes of all the
H($i$)-pairs differ only by a phase for a given qubit $i$.
This state can
always be written in the following way:
\begin{eqnarray}
& |\psi \rangle = a_1|0 \ldots 0^i \ldots  \rangle + a_1e^{i \delta_1}|0 \ldots 1^i \ldots  \rangle \nonumber \\
& + \cdots + a_{2^{n-1}}|1 \ldots 0^i \ldots  \rangle +
a_{2^{n-1}}e^{i \delta_{2^{n-1}}}|1 \ldots 1^i \ldots  \rangle \ .
\label{majlemma1}
\end{eqnarray}
This expression makes explicit that the amplitude for every pair of states
that can be mixed by a Hadamard transformation on the qubit $i$ only
differ by a phase. The Hadamard gate $U_{H,i}$ will mix all these
pairs. The two states in every pair
are equal in all their qubits except the $i$-th one. After
the application of the $U_{H,i}$ we have
\begin{eqnarray}
& U_{H,i}|\psi \rangle = 2^{-1/2} ( a_1(1+e^{i\delta_1})|0 \ldots 0^i
\ldots  \rangle + a_1(1-e^{i \delta_1})|0 \ldots 1^i \ldots  \rangle \nonumber \\
& +  \cdots + a_{2^{n-1}}(1+e^{i\delta_{2^{n-1}}})|1 \ldots 0^i
\ldots \rangle + a_{2^{n-1}}(1-e^{i \delta_{2^{n-1}}})|1 \ldots
1^i \ldots \rangle ) \ .
\label{majlemma2}
\end{eqnarray}
We have to find a set
of probabilities $p_j$ and permutation matrices $P_j$  such that
\begin{equation}
\textstyle \begin{pmatrix}
  |a_1|^2 \\
  |a_1|^2 \\
   \vdots \\
  |a_{2^n-1}|^2 \\
  |a_{2^n-1}|^2 \\
\end{pmatrix}
=\sum_j p_j P_j
\begin{pmatrix}
 |a_1|^2(1+\cos\delta_1) \\
  |a_1|^2(1-\cos\delta_1) \\
   \vdots \\
  |a_{2^n-1}|^2(1+\cos\delta_{2^n-1}) \\
  |a_{2^n-1}|^2 (1-\cos\delta_{2^n-1})\\
\end{pmatrix} \ ,
\label{majlemma3}
\end{equation}
and the unique solution to this probabilistic mixture is
\begin{eqnarray}
&  p_1 = p_2 = \frac{1}{2} \nonumber \\
& \textstyle P_1 =\begin{pmatrix}

  1 &        &          &          &     \\
    & 1      &          &          &     \\
    &        &   \ddots &          &     \\
    &        &          &  1       &     \\
    &        &          &          & 1   \\

\end{pmatrix}
;  \quad P_2 =
\begin{pmatrix}
 0 & 1 &        &   &   \\
 1 & 0 &        &   &   \\
   &   & \ddots &   &   \\
   &   &        & 0 & 1 \\
   &   &        & 1 & 0 \\
\end{pmatrix} \ .
\label{majlemma4}
\end{eqnarray}
The permutation matrix
 $P_1$ is nothing but the identity
matrix and $P_2$ is a permutation of the probabilities of each
pair which has undergone Hadamard mixing. This completes out the
proof. $\square$

\bigskip

 Note that we have made use of the
majorization's definition given in eq. (\ref{maj1}) in this proof.
Consequently, it has not been necessary to introduce any notion of
sorting of the components of the probability distribution. Such a
requirement would have involved an explicit knowledge of the
factors $a_i$ and $\delta_i$ for all $i$, which is unknown to us
in principle.

The lemma we have proven states that Hadamard transformation do
order probability distributions when the input state has a special
structure, namely those amplitudes to be mixed only differ by a
phase. This is the key element pervading the whole proof. Hadamard
transformations and controlled-phase transformations carefully
preserve such a structure when needed as we shall now see.

\subsection{A lemma concerning phase preservation}

\

\textbf{Phase preserving Lemma}

\emph{Let us
consider  a set of Hadamard gates \{$U_{H,j}$\} with $j = 0, 1 \ldots
n-1$ such that each of them can act only once and a quantum state
$|\psi \rangle$ such that the probability amplitudes of the
computational H($j$)-pairs differ only by a phase
$\forall j$.}

\emph{Then, after applying any Hadamard $U_{H,i}$ operator
from this  set, the resultant state retains the property that the
H($j$)-pairs differ only by a phase $\forall j \neq i$.}

\vspace{16pt}

This lemma states the fact the QFT works in such a way that states
to be mixed by Hadamard transformations only differ by a phase all
along the computation, till the very moment when  the Hadamard
operator acts on it. In other words, the structure of gates is
respectful with the relative weights of the H($i$)-pairs.

To prove the above lemma we need to build some intuition.
Let us first go into an example.

We start by introducing a new notation for the phases appearing in
the source quantum state of eq. (\ref{steptres}) to be operated by the
QFT operator. Let us define $ \beta_x \equiv -2 \pi x \phi $.
Then
\begin{equation}
|\psi^{(iii)} \rangle = 2^{-n/2}\sum_{x=0}^{2^n-1}
e^{i\beta_x}|x\rangle \ .
\label{newsteptres}
\end{equation}
Note that, because $x = \sum_{i = 0}^{n-1} x_i 2^i$, we have that
\begin{equation}
\beta_x = \sum_{i=0}^{n-1} -2 \pi x_i 2^i \phi \equiv
\sum_{i=0}^{n-1}x_i \alpha_i \ ,
\label{beta}
\end{equation}
where $\alpha_i \equiv -2 \pi 2^i \phi $.
As an example of this notation, let us write the state in the case
of three qubits:
\begin{eqnarray}
|\psi \rangle =& \frac{1}{2^{3/2}} \left(|000 \rangle + e^{i
 \alpha_2}|100
 \rangle + e^{i \alpha_1}|010 \rangle +  e^{i(\alpha_2+ \alpha_1)}|110 \rangle \right) \nonumber \\
& \quad +\frac{1}{2^{3/2}}\left(|001 \rangle + e^{i \alpha_2}|101
 \rangle
+ e^{i \alpha_1}|011 \rangle +  e^{i(\alpha_2 + \alpha_1)}|111
 \rangle \right)e^{i \alpha_0} \ .
\label{example1}
\end{eqnarray}
We have factorised the $\alpha_0$ phase in the second
line. Alternatively, we can choose to factorise $\alpha_1$
\begin{eqnarray}
|\psi \rangle =& \frac{1}{2^{3/2}} \left(|000 \rangle + e^{i \alpha_2}|100 \rangle + e^{i \alpha_0}|001 \rangle +  e^{i(\alpha_2 + \alpha_0)}|101 \rangle \right) \nonumber \\
& \quad+ \frac{1}{2^{3/2}}\left(|010 \rangle + e^{i \alpha_2}|110
\rangle + e^{i \alpha_0}|011 \rangle +  e^{i(\alpha_2 +
\alpha_0)}|111 \rangle \right)e^{i \alpha_1} \ ,
\label{example2}
\end{eqnarray}
or $\alpha_2$,
\begin{eqnarray}
|\psi \rangle = &\frac{1}{2^{3/2}} \left(|000 \rangle + e^{i \alpha_1}|010 \rangle + e^{i \alpha_0}|001 \rangle +  e^{i(\alpha_1 + \alpha_0)}|011 \rangle \right) \nonumber \\
&\quad + \frac{1}{2^{3/2}} \left(|100 \rangle + e^{i \alpha_1}|110
\rangle + e^{i \alpha_0}|101 \rangle +  e^{i(\alpha_1 +
\alpha_0)}|111 \rangle \right)e^{i \alpha_2} \ ,
\label{example3}
\end{eqnarray}
In total, the initial state for three qubits can be factorised in
 this three different ways.
This example shows that there are three different ways
 of writing the quantum state by focusing on a particular
qubit.
 This property is easily extrapolated to the general case of
$n$-qubits: we can always write the quantum state $|\psi^{(iii)} \rangle$ in
$n$ different ways factorising a particular phase in the second
line.

\vspace{30pt}

{\bf Proof}

In the general case we can factorise the
$\alpha_j$ phase so that the pure state is
written as
\begin{eqnarray}
|\psi \rangle =|\psi^{(iii)}\rangle
 = &\frac{1}{2^{n/2}} \left( |0\ldots 0^{j} \ldots  \rangle + \cdots + e^{i\sum_{k \neq j} \alpha_k}|1\ldots 0^{j} \ldots  \rangle \right) \nonumber \\
& \quad + \frac{1}{2^{n/2}} \left( |0\ldots 1^{j} \ldots  \rangle
+ \cdots + e^{i\sum_{k \neq j} \alpha_k}|1\ldots 1^{j} \ldots
\rangle \right)e^{i\alpha_j} \ .
\label{general}
\end{eqnarray}
Then, the action of the $U_{H,j}$ transforms the state as
follows
\begin{eqnarray}
 U_{H,j}|\psi \rangle = & \frac{(1+e^{i \alpha_{j}})}{2^{(n+1)/2}}
\left( |0 \ldots 0^j \ldots \rangle + \cdots +
e^{i\sum_{k\neq j} \alpha_k}|1 \ldots 0^j
\ldots \rangle \right)
 \nonumber \\
& +\frac{(1-e^{i
\alpha_{j}})}{2^{(n+1)/2}} \left( |0 \ldots 1^j \ldots \rangle + \cdots + e^{i\sum_{k\neq j} \alpha_k} |1  \ldots 1^j \ldots
\rangle \right).
\end{eqnarray}
It is now clear that this resulting state still preserves the
symmetry property necessary to apply the phase preserving lemma to
the  rest of qubits $i\not= j$. The reason is that the effect of the operator has been
splitting the quantum state in two pieces which individually retain the property that all
the H($i$)-pairs differ only by a phase for $i\not= j$. If we now
apply another Hadamard operator over a different qubit $i$,  each of
these two quantum states splits in turn in two pieces
\begin{eqnarray}
 &U_{H,i} U_{H,j}|\psi \rangle = \nonumber \\
&  \frac{(1+e^{i \alpha_{j}})(1+e^{i \alpha_{i}})}{2^{(n+2)/2}}
 \big( |0 \ldots 0^i \ldots 0^j \ldots \rangle
 + \cdots + e^{i\sum_{k\neq i,j} \alpha_k}|1 \ldots 0^i \ldots 0^j
\ldots \rangle \big)
 \nonumber \\
& + \frac{(1+e^{i \alpha_{j}})(1-e^{i \alpha_{i}})}{2^{(n+2)/2}}
 \big( |0 \ldots 1^i \ldots 0^j \ldots \rangle
 + \cdots + e^{i\sum_{k\neq i,j} \alpha_k}|1 \ldots 1^i \ldots 0^j
\ldots \rangle \big)
 \nonumber \\
& + \frac{(1-e^{i \alpha_{j}})(1+e^{i \alpha_{i}})}{2^{(n+2)/2}}
 \big( |0 \ldots 0^i \ldots 1^j \ldots \rangle
 + \cdots + e^{i\sum_{k\neq i,j} \alpha_k}|1 \ldots 0^i \ldots 1^j
\ldots \rangle \big)
 \nonumber \\
& + \frac{(1-e^{i \alpha_{j}})(1-e^{i \alpha_{i}})}{2^{(n+2)/2}}
 \big( |0 \ldots 1^i \ldots 1^j \ldots \rangle
 + \cdots + e^{i\sum_{k\neq i,j} \alpha_k}|1 \ldots 1^i \ldots 1^j
\ldots \rangle \big)
\label{sgon}
\end{eqnarray}
(where we have assumed $i > j$). The register now consists of
a superposition of four quantum states, each made of amplitudes
that only differ by a phase.
Further application of a  Hadamard gate over yet a different qubit
would split each of the four states again in two pieces in a way that the
symmetry would again be preserved. This splitting takes place each
 time a particular Hadamard acts. Thus, all Hadamard gates operate
 in turn producing majorization while not spoiling the symmetry
property needed for the next step. This completes the proof of the
 phase preserving lemma. $\square$

\subsection{ Analysis of the controlled-phase operators}

It is still necessary to verify that the action of
controlled-phase gates does not interfere with the majorization
action carried by the Hadamard gates. Let us concentrate on the
action of $U_{H,n-1}$,
 which is the first Hadamard operator applied in the
 canonical decomposition of the QFT. Originally we had
\begin{eqnarray}
|\psi \rangle = &\frac{1}{2^{n/2}} \left( |0 0 \ldots
 \rangle + \cdots + e^{i\sum_{k \neq n-1} \alpha_k}|0 1 \ldots  \rangle \right)  \nonumber \\
& \qquad +  \frac{1}{2^{n/2}} \left( |1 0 \ldots  \rangle + \cdots
+ e^{i\sum_{k \neq n-1} \alpha_k}|1 1 \ldots  \rangle
\right)e^{i\alpha_{n-1}}  \ ,
\label{first}
\end{eqnarray}
where we have taken out the $\alpha_{n-1}$ phase factor.
After the action of  $U_{H, n-1}$ we get
\begin{eqnarray}
&  U_{H,n-1}|\psi \rangle  =  \frac{(1+e^{i \alpha_{n-1}})}{2^{(n+1)/2}}
\left(|0 0\ldots \rangle + \cdots + e^{i\sum_{k\neq n-1} \alpha_k}|0 1 \ldots \rangle \right) \nonumber \\
& +\frac{ (1-e^{i \alpha_{n-1}})}{2^{(n+1)/2}} \left(|1 0 \ldots \rangle + \cdots + e^{i\sum_{k\neq n-1} \alpha_k} |1 1 \ldots  \rangle
\right)  \equiv  |a\rangle + |b\rangle \ .
\label{second}
\end{eqnarray}
We repeat the observation made in the second step that the  state resulting
from the action of
 $U_{H, n-1}$
 can be divided into a sum of two states, which we
have called $|a \rangle$  and $|b\rangle$. For both of these two
states the amplitudes of the $H(j)$-pairs
$\forall j \neq n-1$ still differ only by a phase.

We can now analyse the effect of the controlled-phase operators.
Notice, first, that any two controlled-phase operators acting over
the same qubit commute, no matter what the phases and the control
qubits are. This is a direct consequence of the diagonal form of
these type of operators, written in the computational basis (see
for example \cite{phase}). Consequently, we can  focus on what
happens after applying a general controlled-phase operator over a
given qubit. For simplicity we will assume that it will act over
the $(n-1)$-th qubit so it will be applied over the state
$U_{H,n-1}|\psi\rangle$ (the following procedure is easily
extrapolated to the controlled-phase operators acting over the
rest of the qubits). If the control qubit is the $l$-th one ($l
\neq n-1$), the operator will only add phases over those
computational states of (\ref{second}) such that both the $1$-th
and the $l$-th qubits are equal to $1$, so we see that it will
only act effectively over part of the $|b\rangle$ state. Let us
write $|b\rangle$ factorising the $l$-th phase
\begin{eqnarray}
 |b \rangle = & \frac{(1-e^{i\alpha_{n-1}})}{2^{(n+1)/2}} \left(
|1\ldots 0^l \ldots  \rangle + \cdots +
e^{i\sum_{k\neq l, n-1}\alpha_k}|1\ldots 0^l
\ldots \rangle \right) \nonumber \\
 & + \frac{ (1-e^{i\alpha_{n-1}})}{2^{(n+1)/2}}
\left(|1\ldots 1^l \ldots  \rangle + \cdots
+e^{i\sum_{k\neq l, n-1}\alpha_k}|1\ldots
1^l \ldots \rangle \right)e^{i\alpha_l} \ .
\label{de}
\end{eqnarray}
It is now clear  that we have only added a global phase in the
second piece of $|b\rangle$, which can always be absorbed in a
redefinition of the phase $\alpha_l$. Hence we see that no
relevant change is made in the quantum state concerning
majorization, because the amplitudes of the computational
$H(j)$-mixable states $\forall j \neq n-1$  still differ only by a
single phase. \emph{The action of controlled-phase operators only
amounts to a redefinition of  phases, which does not affect the
necessary property for the majorization
 lemma to hold}. We see
that the needed phase redefinition can be easily made each time one of
these operators acts over a particular qubit.

\subsection{Summary of the proof}

We can now collect the three pieces of our argument:

\begin{itemize}

\item{Each Hadamard operation acting on a suitably symmetric state
produces majorization.}

\item{ Each time a Hadamard operator is applied, the state splits 
into two states preserving the key property of the majorization 
lemma with respect to the remaining Hadamard operators.}

\item{ Controlled-phase operators only involve phase redefinitions, which are immaterial to majorization.}

\end{itemize}

It then follows the  property that QFT operator majorizes
step by step the probability distribution in phase-estimation
algorithms when the initial state is a set of states differing by
phases. $\square$

\bigskip

We would like to emphasise some relevant points emerging from our
proof. Controlled-phase operators play no role on majorization,
 though they provide entanglement. Curiously, Hadamard operators act
exactly in the complementary way, providing majorization
without providing entanglement.
It is also interesting to note  that the majorization
arrow in the quantum algorithm is based
on two ingredients. On the one hand we have the special
 properties of the quantum state, and on the other hand we have
 the structure of the QFT. A QFT acting on an arbitrary state
would fail to obey majorization. In particular a subsequent
application of a QFT on the final state obtained above
would operate a minorization of probabilities, till reaching
the original state.

It is arguable that the proof we have presented depends on
the specific decomposition of the QFT in terms of individual gates.
The underlying quantum circuit is not unique and majorization may not
be present in alternative decompositions. To clarify this point we 
have analyzed the decomposition of the QFT proposed
in Ref. \cite{moore}. This circuit works in a different
way, implementing first a series of gates that do not change at
all the probability distribution and then a set of Hadamard
operations that do verify step-by-step majorization.
The result remains true that an alternative efficient QFT
obeys step-by-step majorization.

\section{Natural majorization and efficient  quantum algorithms}

We now turn to investigate further the way majorization has
emerged in the phase-estimation algorithm as compared to
majorization in Grover's algorithm.
This comparison was not performed in \cite{lat}.
We shall see that both
algorithms work in a rather different manner. Majorization is more
``natural'' in the former than in the latter. This naturalness is
attached to the absence or presence of  off-diagonal contributions
appearing along the unitary evolution.

For a search in an unstructured database of a particular item, the
best known classical algorithm (which is simply the examination
one by one of all the items in the database) takes asymptotically
${\cal O}(2^n)$ steps in succeeding (where $2^n \equiv N$ is the
number of entries)\cite{grov}. However, Grover was able to find a
quantum mechanical algorithm that implements a quadratic speed-up
versus the classical one (that is, Grover's quantum algorithm
takes asymptotically ${\cal O}(2^{n/2})$ steps). We will not enter
into precise details of the construction of this quantum
algorithm, and will only make few comments on the way it proceeds.

 Grover's algorithm \cite{grov} can be reduced to a two-dimensional Hilbert
 space
 spanned by the state we are searching $|m\rangle$
 and its orthogonal $|m^{\bot}\rangle$\cite{boyer}.
 The unitary evolution of the
quantum state is given by the repeated application of a
kernel which amounts to a rotation
\begin{equation}
\textstyle K =
\begin{pmatrix}
 \cos\theta &  -\sin\theta   \\
 \sin\theta &  \cos\theta   \\
\end{pmatrix}
\label{kernel}
\end{equation}
where $\cos\theta = 1-2/2^n$.
Other choices of kernels are also valid but
the one here presented is the optimal one \cite{mart}. The
initial state of the computation is an equal superposition of all
the computational states, written as $|\psi\rangle =
2^{-n/2}|m\rangle + \left( 1-2^{-n} \right)^{1/2}|m^{\bot}\rangle$
in this 2-dimensional notation. For a given intermediate computation
step the state  $(\alpha, \beta)$ will be transformed to
 $(\alpha', \beta')$. If we wish to express the
initial amplitudes in terms of the final ones, we have:
\begin{equation}
\textstyle
\begin{pmatrix}
 \alpha   \\
 \beta   \\
\end{pmatrix}
=
\begin{pmatrix}
\ \ \ \alpha'\cos\theta + \beta'\sin\theta \\
 -\alpha'\sin\theta + \beta'\cos\theta \\
\end{pmatrix} \ .
\label{evol}
\end{equation}
We now take the modulus squared of the amplitudes, obtaining:
\begin{eqnarray}
& |\alpha|^2 = \cos^2\theta \ |\alpha'|^2 + \sin^2\theta \ |\beta'|^2
 + 2\cos\theta \sin\theta \ {\mathcal Re}(\alpha^{\prime *} \beta')  \nonumber \\
& |\beta|^2 = \sin^2\theta \ |\alpha'|^2 + \cos^2\theta \ |\beta'|^2 -
2\cos\theta \sin\theta \ {\mathcal Re}(\alpha^{\prime *} \beta') \ .
 \label{relations}
\end{eqnarray}
It is clear  that
if the interference terms were to vanish then
 majorization would follow in a straightforward way from these two
relations. To see this, note that in such a case we could obtain
the initial probability set from the final one according to the
definition given in eq. (\ref{maj1}) of majorization, with
probabilities $ p_1 = \cos^2\theta$, $p_2 = \sin^2\theta$ and
permutation matrices $P_1 = I$ and $P_2 = \sigma_1$ (the last one
is just the permutation matrix of the two elements). Majorization
would then follow
 in a natural way  from just
unitary evolution.
But this is not the case in Grover's algorithm,
because interference terms do not vanish. Yet
it is proven that majorization in Grover's algorithm exists
\cite{lat}, although the way it arises is not so directly
related to
the unitary evolution in the way  presented here.

Let us turn back to the majorization in the phase-estimation
algorithm and its relation to unitary evolution. We write the
initial state to be operated upon by a Hadamard gate acting over
the $j$-th qubit as

\begin{eqnarray}
& |\psi \rangle = c_0|0 \ldots 0^j \ldots  \rangle + c_j|0 \ldots 1^j \ldots   \rangle \nonumber \\
& + \cdots + c_{2^{n}-1-j}|1 \ldots 0^j \ldots  \rangle +
c_{2^{n}-1}|1 \ldots 1^j \ldots  \rangle \ ,
\label{estinicial}
\end{eqnarray}
where we are focusing on the  $H(j)$-mixable coefficients. Applying
the Hadamard gate over the $j$-th qubit we get

\begin{eqnarray}
& U_{H,j} |\psi\rangle = 2^{-1/2}  (c_0 + c_j)|0 \ldots 0^j \ldots  \rangle + 2^{-1/2}(c_0 - c_j)|0 \ldots 1^j \ldots  \rangle  \nonumber \\
& + \cdots + 2^{-1/2}(c_{2^{n}-1-j}+c_{2^n-1})|1 \ldots 0^j \ldots  \rangle \nonumber \\
& + 2^{-1/2}(c_{2^n-1-j} - c_{2^{n}-1})|1 \ldots 1^j \ldots  \rangle  \ .
\label{estfinal}
\end{eqnarray}

Inverting the relations in this last equation, we can find the initial probability amplitudes in terms of the final ones. For a given pair of amplitudes $c_{m-j}$ and $c_m$ we find
\begin{eqnarray}
& c_{m-j} = 2^{-1/2}\left(2^{-1/2} \left(c_{m-j}+c_m \right) + 2^{-1/2}\left(c_{m-j}-c_m \right) \right) \nonumber \\
& c_{m} = 2^{-1/2}\left(2^{-1/2} \left(c_{m-j}+c_m \right) -
2^{-1/2}\left(c_{m-j}-c_m \right) \right) \ ,
\label{relamp}
\end{eqnarray}
and making the square modulus we have
\begin{eqnarray}
& |c_{m-j}|^2 = \frac{1}{2} \left( \left| 2^{-1/2} \left( c_{m-j}
+ c_m \right) \right|^2 \right) + \frac{1}{2} \left( \left|
2^{-1/2} \left( c_{m-j} - c_m \right) \right|^2 \right) \nonumber \\
& + \frac{1}{2} {\mathcal Re} \left( (c_{m-j} + c_m)^*
(c_{m-j}-c_m) \right) \nonumber \\
& |c_{m}|^2 = \frac{1}{2} \left( \left| 2^{-1/2} \left( c_{m-j} + c_m \right) \right|^2 \right) + \frac{1}{2} \left( \left| 2^{-1/2} \left( c_{m-j} - c_m
\right) \right|^2 \right) \nonumber \\
& - \frac{1}{2} {\mathcal
Re} \left( (c_{m-j} + c_m)^* (c_{m-j}-c_m) \right)  \ .
\label{relmodul}
\end{eqnarray}

As in the Grover's previous example, we observe that if
interference terms disappeared majorization would
trivially arise from this
set of relations. In such a case, we  would only have to choose the
set of probabilities and permutation matrices of (\ref{majlemma4})
to prove this property. But again we note that in general those
interference terms do not disappear, so majorization does not
usually arise in this natural way from the unitary evolution
(exactly the same that happened in Grover's algorithm). For those
terms to disappear, there must exist very specific properties
for the coefficient $c_i$. It
is a straightforward exercise checking that the interference
vanish if and only if
\begin{eqnarray}
& c_{m-j} = a_{m-j} \nonumber \\
& c_m = a_{m-j}e^{i \delta_{m-j}}  \ ,
\label{ifandonlyif}
\end{eqnarray}
where $a_{m-j}$ is real.

It is a remarkable  fact that this is the case of phase-estimation
algorithms. Recalling the lemmas from the previous sections, the
interference terms will vanish also step-by-step, so \emph{in the
case of phase-estimation algorithms, step-by-step majorization
arises as a natural consequence of unitary evolution}, something
that does not happen in a general case. We also realize that
\emph{the form of the quantum state to be applied to the QFT
 in eq. }(\ref{steptres}) \emph{is precisely the unique possible form for
step-by-step natural majorization to appear}. In a way we can say
that previous steps in the algorithm prepare the state in this
unique form.

Actually the test for natural majorization
 can be performed for any unitary operation
 acting over any quantum state and consequently for \emph{any}
 quantum algorithm.
 To make this observation clear, let us write in matrix notation a
unitary evolution of a general quantum state in an $N$-dimensional  Hilbert space:
\begin{equation}
\begin{pmatrix}
\alpha'_1 \\
\vdots \\
\alpha'_N \\
\end{pmatrix}
= U
\begin{pmatrix}
\alpha_1 \\
\vdots \\
\alpha_N \\
\end{pmatrix} \ ,
\label{evoluc}
\end{equation}
where $\alpha_i$ are the amplitudes for the initial state, $\alpha'_i$ the amplitudes for the final state, and $U$ corresponds to the $N \times N$ matrix of a general unitary evolution. Inverting this relation we can obtain the original amplitudes in terms of the final ones. Now, if we write $U^{\dag}$ as
\begin{equation}
U^{\dag} =
\begin{pmatrix}
u_{11} & \cdots & u_{1N} \\
\vdots & \ddots & \vdots \\
u_{N1} & \cdots & u_{NN} \\
\end{pmatrix}
\label{unitaria}
\end{equation}
we can again calculate the square modulus of the original amplitudes obtaining
\begin{eqnarray}
& |\alpha_1|^2 = |u_{11}|^2|\alpha'_1|^2 + \cdots +
|u_{1N}|^2|\alpha'_N|^2 + {\rm interference \ terms} \nonumber \\
& \vdots  \nonumber \\
& |\alpha_N|^2 = |u_{N1}|^2|\alpha'_1|^2 + \cdots +
|u_{NN}|^2|\alpha'_N|^2 + {\rm interference \ terms} \ .
\label{relations2}
\end{eqnarray}
We see that if the interference terms vanish, we find:
\begin{equation}
\begin{pmatrix}
|\alpha_1|^2 \\
\vdots \\
|\alpha_N|^2 \\
\end{pmatrix}
= D
\begin{pmatrix}
|\alpha'_1|^2 \\
\vdots \\
|\alpha'_N|^2 \\
\end{pmatrix}
\label{major}
\end{equation}
where $D$ is a doubly stochastic matrix such that $D_{ij} \equiv
|u_{ij}|^2$. Recalling eq. (\ref{maj3}) we conclude that the final
probability distribution majorizes the original one, in a way that
majorization arises in a natural way from the unitary evolution.
This result is reminiscent of other applications of majorization
theory in quantum mechanics. For instance, ensemble realization
is rooted in doubly stochastic matrices of the above form: given a density matrix
 $\rho$ and a probability distribution $p_i$, there exist normalised quantum states
$|\psi_i\rangle$ such that
\begin{equation}
\rho=\sum_i p_i |\psi_i\rangle \langle \psi_i|
\label{dens}
\end{equation}
if and only if $(p_i) = D \lambda_{\rho}$, where $\lambda_{\rho}$ is the vector of eigenvalues of $\rho$ and $D_{ij} \equiv |u_{ij}|^2$, being $u_{ij}$ the elements of a unitary matrix \cite{vid}.

\section{Conclusions}

Our main results in this paper have been to provide
an explicit and detailed  proof that the
Quantum Fourier Transformation implements 
 step by step majorization in phase-estimation quantum
algorithms, along with a comparison between the different way
in which the notion of majorization \cite{lat}
operates for the two families of efficient quantum algorithms known so far.
Further work on other quantum algorithms
(Bernstein-Vazirani problem, determinaiton of parity,
quantum adiabatic evolution and quantum random walks for
non-trivial graphs) that will be presented
elsewhere \cite{inpreparation} reinforces the link between
step-by-step majorization and efficiency.

The proof of majorization for QFT is based on introducing the
concept of Hadamard pair, H($i$)-pair, and three
lemmas. A H($i$)-pair corresponds to any pair of states which
can be mixed by a Hadamard transformation on the bit $i$.
The three lemmas show that:
\begin{itemize}
\item A Hadamard gate acting on the bit $i$ of a given
register  carries out majorization provided
the register is made of H($i$)-pairs where
the two states in the pair only differ by a phase.
\item A Hadamard gate on a bit $i$  preserves the
relative phase structure of
H($j$)-pairs for $j\not= i$.
\item Controlled-phase gates do not affect the phase structure of
any H($i$)-pair.
\end{itemize}

Step-by-step majorization then follows
since the Quantum Fourier Transformation only uses one Hadamard gate
per qubit supplemented with immaterial controlled-phase gates.
The
detailed way in which majorization takes place suggest introducing the
concept of {\sl natural majorization}, defined as a majorization
which follows from unitary evolution in the absence of
off-diagonal terms. Natural majorization controls phase-estimation
algorithms but is not present in Grover's algorithm. A
classification of quantum algorithms according to a naturalness
criterium for majorization can be performed. We note that this
classification  furthermore
 corresponds to the well classification in terms of
their efficiency. Natural majorization might be a distinct feature
of exponential quantum speed up.

\textbf{Acknowledgements:} we are grateful to
Enric Jan\'e, Llu\'{\i}s Masanes, Enrique Rico  and Guifr\'e Vidal
for discussions
about the content of this paper. Financial support has been obtained from the
projects AEN99-0766, 19999GR-00097, IST-1999-11053, PB98-0685
and  BFM2000-1320-C02-01.

\end{document}